\documentclass[conference, 10pt]{IEEEtran}
\usepackage{ifthen}
\usepackage[utf8]{inputenc}
\usepackage{latexsym, amssymb, dsfont}
\usepackage{amsmath,amsthm}
\usepackage{amssymb}
\usepackage[inline]{enumitem}
\usepackage{siunitx}
\usepackage{graphicx}
\usepackage{caption}
\usepackage{todonotes}
\usepackage[caption=false,font=footnotesize]{subfig}
\usepackage{orcidlink}
\usepackage{hyperref}
\usepackage[capitalize]{cleveref}

\usepackage{tikz}

\newcommand{\dupD}{\delta^\uparrow}
\newcommand{\ddoD}{\delta^\downarrow}

\newcommand{\dsu}{\delta_S^\uparrow}
\newcommand{\dsd}{\delta_S^\downarrow}
\newcommand{\dmu}{\delta_M^\uparrow}
\newcommand{\dmd}{\delta_M^\downarrow}

\newcommand{\figPath}[1]{#1}

\newcommand{\NOR}{\texttt{NOR}}

\newcommand{\dmin}{\delta_{\mathrm{min}}}

\newcommand{\spice}{\textit{SPICE}}

\newcommand{\fud}{f_{\uparrow/\downarrow}}

\newcommand{\vth}{V_{th}}

\newcommand{\vdd}{V_{DD}}
\newcommand{\gnd}{\textit{GND}}
\newcommand{\vout}{V_{O}}

\newcommand{\va}{V_{A}}
\newcommand{\vb}{V_{B}}

\newcommand{\vint}[1]{V_{\mathit{N}}^{#1}}
\newcommand{\cint}[1]{C_{\mathit{N}}^{#1}}
\newcommand{\iintm}[1]{I_{\mathit{N}}^{#1}}

\newcommand{\cout}{C_{\mathit{O}}}

\newcommand{\iout}{I_{\mathit{O}}}

\newcommand{\nmos}{nMOS}
\newcommand{\pmos}{pMOS}

\newcommand{\ohm}{(OHM)}

\definecolor{color1}{HTML}{E41A1C}
\definecolor{color2}{HTML}{377EB8}
\definecolor{color3}{HTML}{4DAF4A}
\definecolor{color4}{HTML}{984EA3}
\definecolor{color5}{HTML}{FF7F00}

\hypersetup{
  colorlinks=true,
  urlcolor=color2,
  citecolor=color3,
  linkcolor=black
}

\newboolean{conference}
\setboolean{conference}{false}  

\crefname{equation}{}{}

\title{A Simple Hybrid Model for Accurate Delay Modeling of a Multi-Input 
  Gate\thanks{This research was partially funded by the Austrian Science Fund (FWF)
    projects DMAC (P32431).}}

\author{
  \IEEEauthorblockN{
    Arman Ferdowsi\,\orcidlink{0000-0002-9374-3828},
    J\"urgen~Maier\,\orcidlink{0000-0002-0965-5746},
    Daniel~\"Ohlinger\,\orcidlink{0000-0001-8097-3619},
    Ulrich~Schmid\,\orcidlink{0000-0001-9831-8583}
  }
  \IEEEauthorblockA{TU Wien, ECS Group (E191-02)\\
    \{aferdowsi, jmaier, doehlinger, s\}{@}ecs.tuwien.ac.at}
}

\IEEEoverridecommandlockouts 
\DeclareUnicodeCharacter{2212}{-}
\begin{document}

\maketitle
	
\begin{abstract}
  Faithfully representing small gate delay variations caused by input
  switchings on different inputs in close temporal proximity is a very
  challenging task for digital delay models. In this paper, we use the example
  of a 2-input NOR gate to show that a simple hybrid model leads to a
  surprisingly accurate digital delay model. Our model utilizes simple
  first-order ordinary differential equations (ODEs) in all modes, resulting
  from considering transistors as ideal switches in a simple RC model of the
  gate.  By analytically solving the resulting ODEs, we derive expressions for
  the gate delays, as well as formulas that facilitate model
  parametrization. It turns out that our model almost faithfully captures the
  Charlie effect, except in just one specific situation. In addition, we
  experimentally compare our model's predictions both to SPICE simulations,
  using some \SI{15}{\nm} technology, and to some existing delay
  models. Our results show a significant improvement of the achievable
  modeling accuracy.

\end{abstract}

\section{Introduction}
\label{sec:intro}

Digital circuit design relies heavily on fast digital timing analysis
techniques, since they are orders of magnitude faster than analog simulations,
e.g., in \spice. In contrast to static approaches, dynamic digital timing
analysis predicts the propagation of arbitrary signal traces throughout a
circuit. Going beyond the popular pure (= constant input-to-output delay) and
inertial delay (= constant delay + too short pulses being removed)
models~\cite{Ung71}, \emph{single-history delay models}~\cite{BJV06,FNNS19:TCAD}
achieve an improved behavioral coverage.  Indeed, as proved in \cite{FNS16:ToC},
it is inevitable for any faithful delay model that a gate's input-to-output
delay $\delta(T)$, for a given transition, depends on a parameter like the
previous-output-to-input delay $T$.

The \emph{involution delay model} (IDM) proposed in \cite{FNNS19:TCAD} consists
of zero-time boolean gates, which are interconnected by single-input
single-output involution delay channels. IDM channels are characterized by a
delay function $\delta(T)$, which is a negative involution, in the sense that
$-\delta(-\delta(T))=T$.  Unlike all other existing delay models, the IDM
faithfully models glitch propagation in the simple short-pulse filtration
problem, and is hence the only candidate for a faithful delay model known so
far.

Moreover, the IDM comes with a publicly available timing analysis framework (the
\emph{Involution Tool}~\cite{OMFS20:INTEGRATION}), which is based on an
industrial simulation suite.  The Involution Tool allows to randomly generate
input traces for a given circuit, and to evaluate the accuracy of IDM
predictions compared to \spice-generated transition times and/or other digital
models like inertial delays.

Whereas the accuracy of IDM predictions for single-input, single-output circuits
like inverter chains or clock trees reported in~\cite{OMFS20:INTEGRATION} is
impressive, this is less so for circuits involving multi-input gates. We
conjecture that this is mainly due to the inherent lack of properly covering
output delay variations caused by \emph{multiple input switching} (MIS) in close
temporal proximity~\cite{CGB01:DAC}, also known as the \emph{Charlie effect}
(named after Charles Molnar, who identified its causes in the 70th of the last
century).  Compared to the \emph{single input switching} (SIS) case, the output
transition is sped up/slowed down with decreasing transition separation time on
different inputs here.  Single-input, single-output IDM delay channels obviously
cannot exhibit such a behavior.

Multiple approaches have been proposed to cover MIS effects in literature,
ranging from linear~\cite{SRC15:TDAE} or quadratic fitting~\cite{SKJPC09:ISOCC}
over higher-dimensional model representation~\cite{SC06:DATE} to recent machine
learning methods~\cite{RS21:TCAD}. However, none of these naturally generalizes
to multi-input involution channels.

In order to define a 2-input IDM channel, we thus generalize the simple analog
first-order model matching a classic IDM channel (which can be viewed as a
hybrid model with only two modes)\footnote{One mode for generating a rising
  transition switching waveform, and one for the falling one.}  to a four-mode
hybrid model: For each state of the inputs
$(A,B)\in \{(0,0), (0,1), (1,0), (1,1)\}$, a system of first-order
\emph{ordinary differential equations} (ODEs) is derived, which governs the
analog trajectory of the gate's output in the respective mode. At an input
change, the mode is instantaneously switched, in a way that guarantees
\emph{continuity} of the output signal.  Whereas similar approaches have been
advocated in~\cite{Melcher92:MAM, AB06:SST}, these rely on analog fitting or
extraction of unique switching waveforms.
	
\paragraph*{Main contributions}
(1)We introduce a simple hybrid ODE model of a 2-input CMOS \NOR\ gate, which
results from replacing transistors in a simple RC model of the circuit by ideal
switches that are switched on/off at the respective input threshold voltage
$\vth=\vdd/2$ crossing times. We analytically solve the ODE systems for every
mode, and derive expressions for the resulting MIS gate delay $\delta(\Delta)$,
defined by the time when the output waveform crosses $\vth$; the parameter
$\Delta=t_B-t_A$ denotes the switching time separation of the inputs.  It turns
our that the resulting delay model captures all MIS effects very well, except
for one case ($\Delta < 0$ for rising output transitions).

(2) We develop expressions to ease the parametrization of our ODE model, given
the characteristic SIS delay values $\delta(-\infty)$ and $\delta(\infty)$ as
well as the MIS value $\delta(0)$. Since it turned out that it is impossible to
simultaneously match all three values for a real circuit simultaneously, we had
to add an additional pure delay (supported by the IDM) to our parametrization.

(3) We use an appropriately extended version of the Involution Tool to compare the
average modeling accuracy of our hybrid model to other analog/digital
simulations. To that end, we evaluate random traces for one of the circuits
studied in~\cite{OMFS20:INTEGRATION}, using some empirically optimal
parametrization of our hybrid model. Whereas the latter outperforms the original
IDM, as well as standard models like inertial delays, its deficiency in fully
capturing all MIS effects also somewhat impairs its average accuracy.

In a nutshell, our results show that describing multi-input gates with hybrid
models is beneficial, albeit our instantaneously switching transistor
abstraction is slightly too simplistic to fully cover all MIS effects.

\begin{figure}[tbp]
  \centering
  \includegraphics[width=.75\linewidth]{\figPath{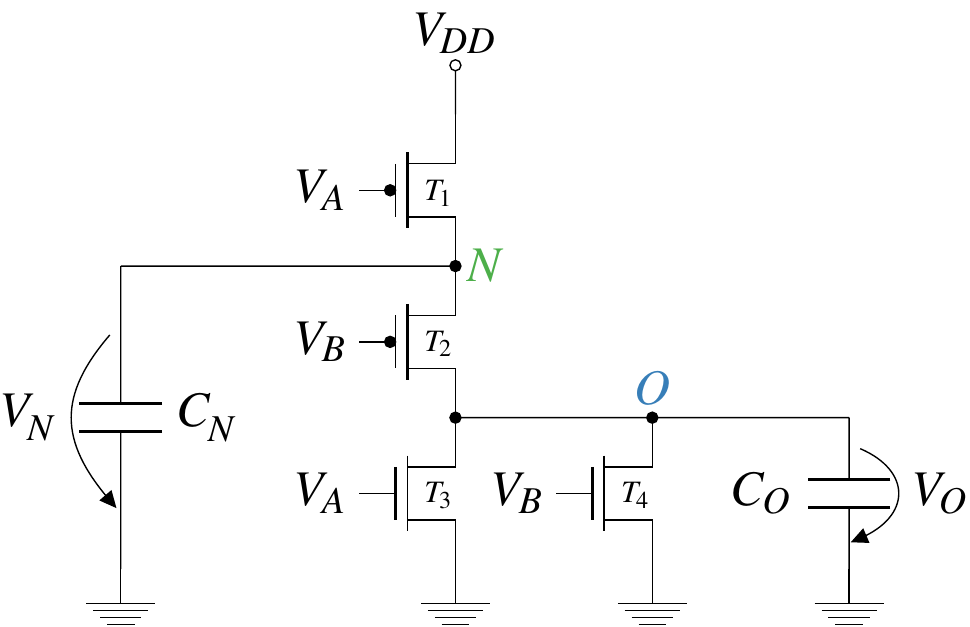}}
  \caption{\label{fig:nor_CMOS} Transistor level implementation of
the \NOR\ gate.}
\end{figure}
	
\paragraph*{Paper organization}
In \cref{sec:SPIC}, we explain the causes for MIS delay variations and determine
characteristic values for a \NOR\ gate.  In \cref{sec:simplehybridmodel}, we
present our simple hybrid ODE model and explore in \cref{sec:charlie} its
ability to capture MIS effects. \cref{sec:param} provides formulas for
parametrizing the model, and \cref{sec:modelingaccuracy} quantifies the average
modeling accuracy.  Some conclusions and directions of future research close the
paper in \cref{sec:conclusions}.

\begin{figure*}[t!]
  \centering
  \subfloat[Falling output transition]{
    \includegraphics[width=.23\linewidth]{\figPath{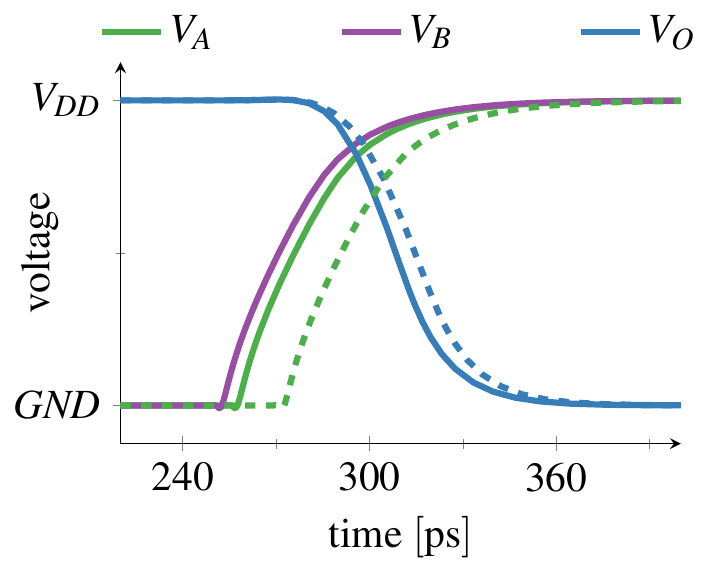}}
    \label{fig:nor2_out_down}}
  \hfil
  \subfloat[Falling output delay]{
    \includegraphics[width=0.23\linewidth]{\figPath{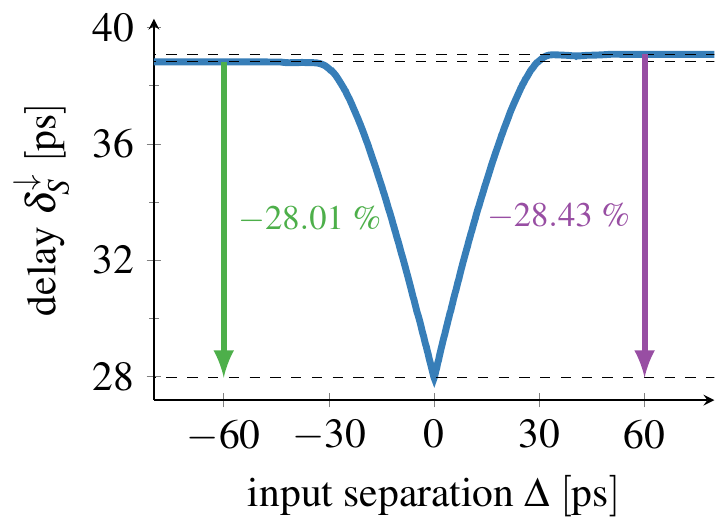}}%
    \label{fig:nor2_out_down_charlie}}
  \hfil
  \subfloat[Rising output transition]{
    \includegraphics[width=.23\linewidth]{\figPath{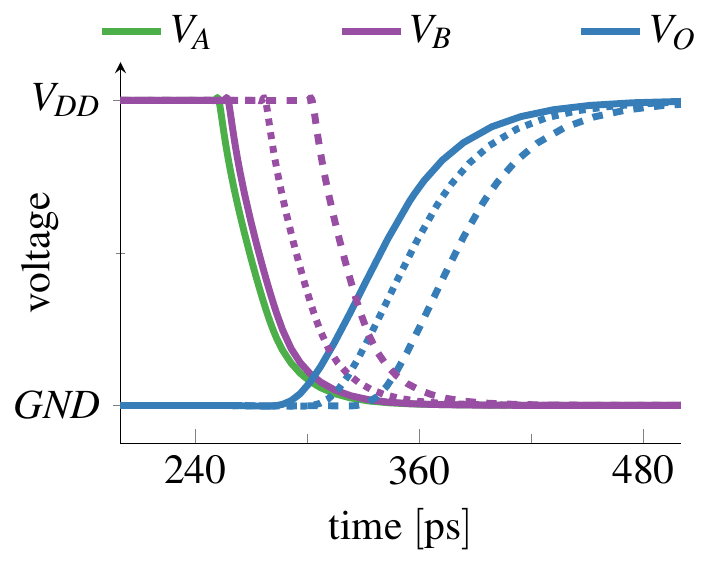}}
    \label{fig:nor2_out_up}}
  \hfil
  \subfloat[Rising output delay]{
    \includegraphics[width=0.23\linewidth]{\figPath{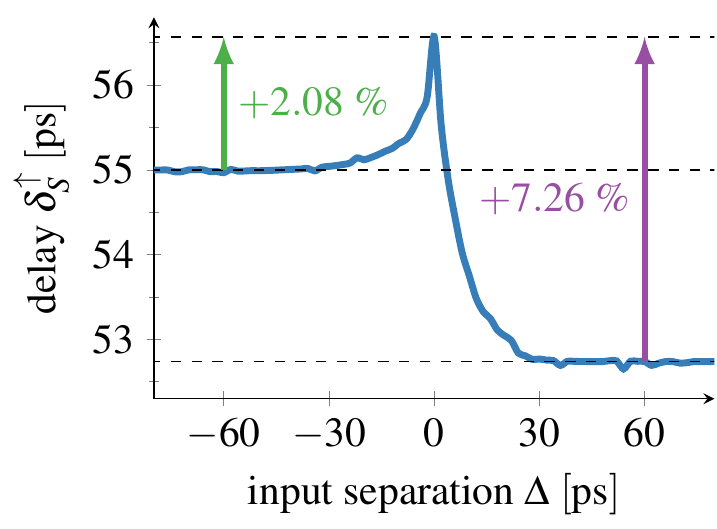}}%
    \label{fig:nor2_out_up_charlie}}
  \caption{Analog simulation results for the CMOS \NOR\ gate.}\label{fig:nor_analog_sim}
\end{figure*}

\section{Multiple Input Switching (MIS)}
\label{sec:SPIC}
	
In this section, we will provide some basic explanations for MIS effects and
quantify those by conducting analog simulations using Spectre (version 19.1) and
the Nangate Open Cell Library featuring FreePDK15$^\text{TM}$ FinFET
models~\cite{Nangate15}. The investigated CMOS \NOR\ gate is among the simplest
multi-input gates and hence a natural target for our analysis. Its
transistor-level implementation, with the parasitic capacitance $\cint{}$ and
the output load capacitance $\cout$, is shown in \cref{fig:nor_CMOS}.  As we
will explore later, the transistor arrangement plays a decisive role in
explaining the observed delay variations: While the \pmos\ ($T_1$ and $T_2$) are
connected in series towards $\vdd$, the \nmos\ ($T_3$ and $T_4$) provide
parallel paths towards $\gnd$.

In the sequel, we apply the rising/falling input waveforms $\fud(t-t_A)$ on
input $A$ resp. $\fud(t-t_B)$ on input $B$, whereat $t_A$ resp. $t_B$ denote the
point in time the \emph{discretization threshold voltage} $\vth=\vdd/2$ is
crossed. In the same spirit, $t_O$ denotes the time when the output voltage
$\vout$ crosses $\vdd/2$. Varying $t_A$ and $t_B$ allows us to represent the
gate delay\ifthenelse{\boolean{conference}} {}{ by $t_O-t_A$ resp. $t_O-t_B$
  (depending on the particular output state)} over the relative input separation
time $\Delta=t_B-t_A$.
	
We first consider the case of a falling output transition. In a
nutshell, either transistor $T_3$ or $T_4$ starts to conduct (is closed), while
one of the two \pmos\ transistors in series stops conducting (is
opened). Consequently, the output is drained and $\vout$ starts to decrease. It
is easy to see that it makes a difference whether only one or both \nmos\
transistors are closed, i.e., only a single or both inputs switch, as it takes,
at least theoretically, only half the time to drain the output in
parallel. So the MIS causes a \textit{speed-up} here, whose effect is clearly
visible in the analog waveforms shown in \cref{fig:nor2_out_down}: When both
transistors start to conduct, the output slope changes notably, leading to a
reduction of the gate delay.

As the first rising input transition already induces an output transition, the
relevant gate delay is $\dsd(\Delta)=t_O - \min(t_A, t_B)$, i.e., the time
difference between the threshold crossing of the output and the earlier input
(see \cref{fig:nor2_out_down_charlie}). As predicted, the delay is the smallest
for simultaneous transitions ($\Delta=0$), whereat the change in delay is around
$30 \%$. Note that $\dsd(-\infty)\neq \dsd(\infty)$ is mainly caused by
transistor $T_2$, which is closed in one case, connecting nodes $N$ and $O$,
while in the other it is open (see \cref{sec:simplehybridmodel}). Although the
absolute values differ, our results fit very well\footnote{We ran our
  simulations with an older technology library (\SI{65}{\nm}) as well, which
  confirmed the delay values reported in the literature.} to previous
investigations in other technologies~\cite{SRC15:TDAE,SKJPC09:ISOCC,SC06:DATE}.

A less visible phenomenon of the speed-up MIS effect deserve to be mentioned
here: Simulations for an older \SI{65}{\nm} technology and results reported in
the literature (e.g. in \cite{SRC15:TDAE,SKJPC09:ISOCC}) reveal local delay
maxima for medium-sized $|\Delta|$. We conjecture these to be caused by
input-output coupling capacitances, which introduce a current working against
the intended behavior of the gate. If the input transitions are far apart
($|\Delta|\gg0$), the second one appears way after $t_O$, such that the
introduced current has no impact. For decreasing $|\Delta|$, however, the second
transition will eventually occur when $\vout$ is about to cross the
threshold. Since the additional current has to be compensated by the driving
transistor, the output slope decreases and thus the delay increases. Further
reducing $\Delta$ first amplifies this effect until the closing of the second
transistor leads to an increased conductivity that, overall, is able to make up
for the added delay, causing it to finally drop again.

For rising output transitions, the behavior of the \NOR\ is quite
different. First and foremost, the gate only switches after both inputs have
changed (see \cref{fig:nor2_out_up}), resulting in the gate delay
$\dsu(\Delta)=t_O - \max(t_A, t_B)$. At the transistor level, each falling input
transition causes one of the \nmos\ to stop conducting while simultaneously one
of the \pmos\ gets closed. We emphasize that the shape of the output
signal in \cref{fig:nor2_out_up} is essentially independent of $\Delta$, only
the position in time varies. This is in accordance with the fact that there is
only a single path connecting the output to $\vdd$.

The SIS delays $\dsu(\infty)$ and $\dsu(-\infty)$ again differ (see
\cref{fig:nor2_out_up_charlie}), i.e., the gate delay depends on the
order of the input transitions. Taking a closer look at the schematics in
\cref{fig:nor_CMOS} reveals that an early transition on $A$ closes the topmost
transistor and thus causes the internal node $N$ between the \pmos\ to be
charged to $\vdd$. By contrast, an early transition on $B$ causes $N$ to be
fully discharged, which obviously prolongs the transition time.
	
In any case, the gate delay for $|\Delta| \to 0$ increases, i.e., the MIS effect
is a \textit{slow-down} here.  The causes are, once again, coupling
capacitances, this time between $N$ and the input: If both inputs switch at the
same time, the parasitic current (dis)charges $\cint{}$, possibly below $\gnd$,
since both adjacent transistors are still open. After they start to conduct, the
additional charge has to be compensated, which explains the increased
delay. Naturally, the delay variations depend on the initial value of $\vint{}$
and thus on the switching history of the gate. Note that we used the worst case
($\vint{}=\gnd$) in all our simulations.

\section{Simple hybrid ODE model}
\label{sec:simplehybridmodel}
	
In our attempt to analytically express the gate delays of a \NOR\ gate, we
replace the transistors by zero-time switches: Depending on whether the
appropriate input is above (logical 1) resp.\ below (logical 0) $\vth=\vdd/2$,
an \nmos\ transistor is replaced by a fixed resistor $R<\infty$ or removed
($R=\infty$), while a \pmos\ is handled in the opposite way. Note that this is
similar to the approach used in~\cite{SHWW14:ICCAD}, with the main difference
that we added a capacitance at the internal node in the p-stack ($\cint{}$) and
one at the output ($\cout$) (cf. \cref{fig:nor_CMOS}). Thus, we end up
with a system of coupled first-order differential equations.

\begin{figure*}[t!]
  \centering
  \subfloat[system $(1,1)$]{
    \includegraphics[width=.23\linewidth]{\figPath{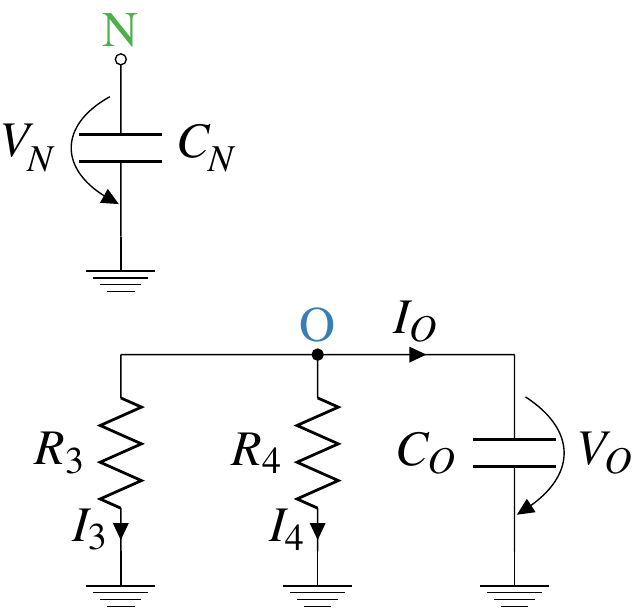}}
    \label{fig:nor_RC_case11}}
  \hfil
  \subfloat[system $(1,0)$]{
    \includegraphics[width=0.23\linewidth]{\figPath{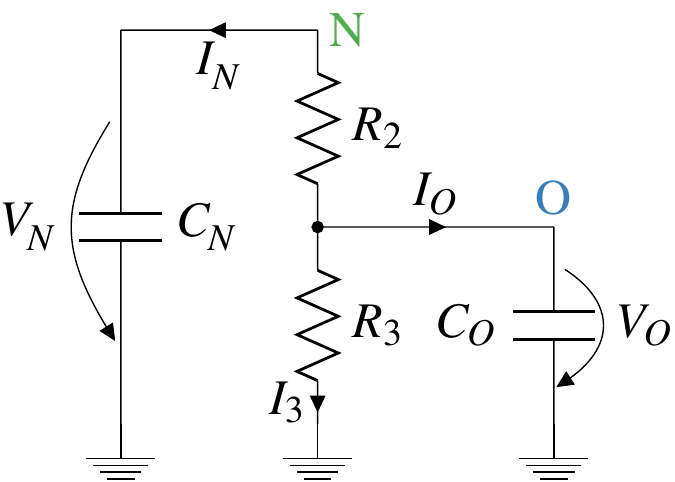}}%
    \label{fig:nor_RC_case10}}
  \hfil
  \subfloat[system $(0,1)$]{
    \includegraphics[width=.23\linewidth]{\figPath{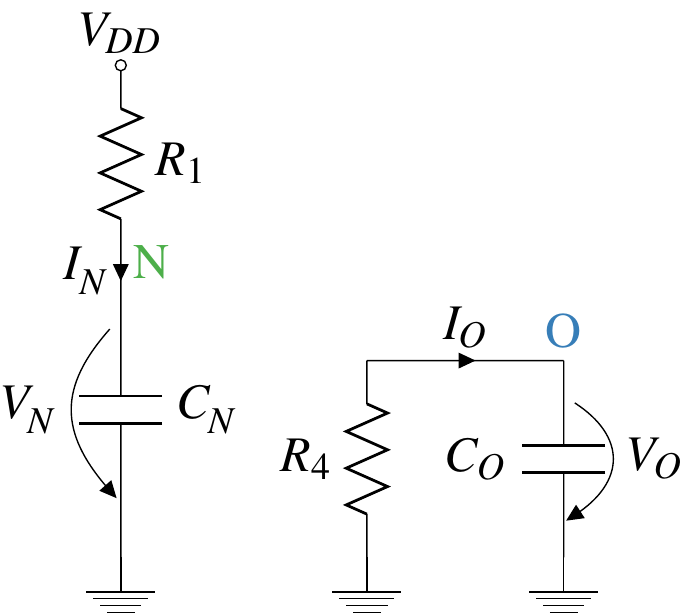}}
    \label{fig:nor_RC_case01}}
  \hfil
  \subfloat[system $(0,0)$]{
    \includegraphics[width=0.14\linewidth]{\figPath{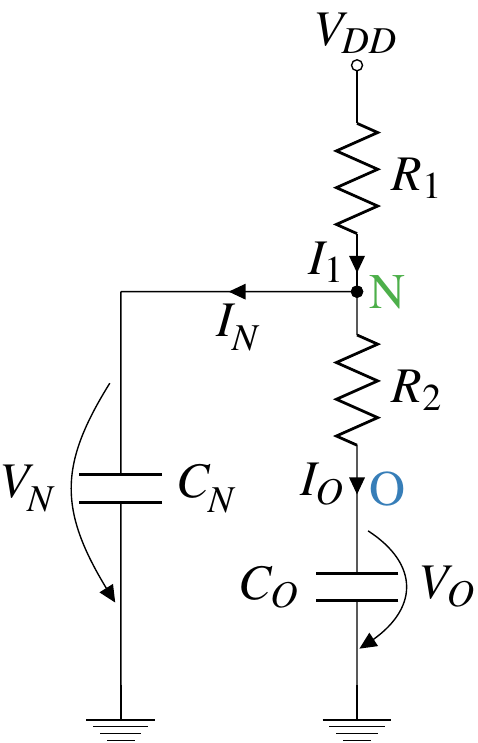}}%
    \label{fig:nor_RC_case00}}
  \caption{First order RC approximations.}\label{fig:modes}
\end{figure*}

\subsection{General solution}

We briefly recall the general solution of first-order ODE systems first.  Let
$V: \mathbb{R} \rightarrow \mathbb{R}^n$ and
$V' = {\operatorname{d}\over\operatorname{d}\!t} V$, and consider a homogeneous
system of ordinary differential equations (ODEs) with constant coefficients
$V'(t) = A \cdot V(t)$; $A=[a_{ij}]^{n \times n} \in \mathbb{R}^{n \times n}$
and initial values $V(0)$. For a diagonizable matrix $A$, this system has a
general solution
\begin{equation*} V(t) = c_1 \cdot \epsilon_{1} \cdot e^{\lambda_{1}t} +
  \ldots + c_n \cdot \epsilon_{n} \cdot e^{\lambda_{n}t} = c\cdot \phi(t)
\end{equation*}
where $ \{ (\lambda_{i}, \epsilon_{i}) \}_{i=1}^{n}$ is a
set of pairs consisting of $n$ eigenvalues and the corresponding eigenvectors of
$A$.  $c= [c_1, \ldots, c_n]^T$ is made up of arbitrary real constants
determined by $V(0)$, and $\phi(t) =[\epsilon_{1} e^{\lambda_1 t}, \ldots,
\epsilon_{n} e^{\lambda_n t}]$ is the fundamental matrix solution of the
homogeneous system. Moreover, the general solution to the non-homogeneous system
$V'(t) = A \cdot V(t) + g(t)$, for a continuous function $g: \mathbb{R}
\rightarrow \mathbb{R}^{n}$, is given by the sum of the general solution of the
corresponding homogeneous system $V(t)' = A \cdot V(t)$ plus a particular
solution to the non-homogeneous one. To be more precise, $V(t) = \phi(t) \cdot c
+ \phi(t) \cdot \int \phi^{-1}(s) \cdot g(s) ds$. We refer the interested reader
to standard textbooks such as \cite{strang2014differential} for more
information.

Since there are four different states $(0,0)$, $(0,1)$, $(1,0)$ and $(1,1)$ of
the inputs $(A,B)$, interpreted as binary signals, we need to consider 4
different RC circuits and their corresponding ODE systems
$V'(t) = A \cdot V(t) + g(t)$. $V(t) \in \mathbb{R}^2$ is a two-element vector,
representing the voltage $\vint{}$ at the internal node $N$ in
\cref{fig:nor_CMOS}, and the gate output voltage $\vout$, i.e.,
\begin{align*}
  V(t)=\left(
  \begin{array}{c}
    \vint{} \\
    \vout \\
  \end{array} \right) \,,
\end{align*}
while the non-homogeneous term $g(t)$ is either identically zero or a constant.
In the sequel, we will evaluate $V(t)$ for each input combination individually.

\subsection{System $(1,1)$: $\va=\vdd$, $\vb=\vdd$}

If inputs $A$ and $B$ are above the threshold, both \nmos\ transistors are
conducting and thus replaced by resistors (see \cref{fig:nor_RC_case11}),
causing the output $O$ to be discharged in parallel. By contrast, $N$ is
completely isolated and keeps its value.  Formally, we obtain

{\small
  \begin{align*}
    \cint{} \cdot {\operatorname{d}\over\operatorname{d}\!t} \vint{} &= 0,\\
    \cout \cdot {\operatorname{d}\over\operatorname{d}\!t} \vout &= \iout = 
                                                                   - I_3 -
                                                                   I_4 = - \vout \cdot \left( \frac{1}{R_3} + \frac{1}{R_4} \right) . 
	\end{align*}
}

This homogeneous system can be rewritten in matrix form
\begin{flalign*}
  &\left(
    \begin{array}{c}
      {\operatorname{d}\over\operatorname{d}\!t} \vint{} \\
      {\operatorname{d}\over\operatorname{d}\!t} \vout \\
    \end{array} \right)
  =
  \left(
    \begin{array}{cc}
      0 & 0 \\
      0 &  - (\frac{1}{\cout R_3} + \frac{1}{\cout R_4}) \\
    \end{array} \right)
  \cdot 
  \left(
    \begin{array}{c}
      \vint{} \\
      \vout \\
    \end{array} \right),&
\end{flalign*}
which leads to the general solution
\begin{flalign*}
  &\left(
    \begin{array}{c}
      \vint{} \\
      \vout \\
    \end{array} \right)
  =c_1
  \left(
    \begin{array}{c}
      1 \\
      0 \\
    \end{array} \right)
  +
  c_2
  \left(
    \begin{array}{c}
      0 \\
      1 \\
    \end{array} \right) \cdot e^{-\bigl(\frac{1}{\cout R_3} + 
    \frac{1}{\cout R_4}\bigr)t}\ .&
\end{flalign*}

\subsection{System $(1,0)$: $\va=\vdd$, $\vb=\gnd$}
	
Since $T_1$ and $T_4$ are open (see \cref{fig:nor_RC_case10}), node $N$ is
connected to $O$, and $O$ to \gnd. Note that both capacitances have to be
discharged over resistor $R_3$, resulting in less current that is available for
discharging $\cout$. More specifically, we observe $\iout=-I_3-\iintm{}$ and
hence obtain

{\small
  \begin{align*}
    \cint{} \cdot {\operatorname{d}\over\operatorname{d}\!t} \vint{} &=  
                                                                       \iintm{} = -
                                                                       \frac{\vint{}-\vout}{R_2}, \\
    \cout \cdot {\operatorname{d}\over\operatorname{d}\!t} \vout &=  \iout 
                                                                   = -
                                                                   I_3 - \iintm{}
                                                                   = - \frac{\vout}{R_3} + \frac{\vint{}-\vout}{R_2}.
  \end{align*}
}
The matrix form of this homogeneous system is
\begin{align*}
  \left(
  \begin{array}{c}
    {\operatorname{d}\over\operatorname{d}\!t} \vint{} \\
    {\operatorname{d}\over\operatorname{d}\!t} \vout \\
  \end{array} \right)
  =&
     \left(
     \begin{array}{cc}
       -\frac{1}{\cint{} R_{2}} & \frac{1}{\cint{} R_{2}} \\
       \frac{1}{\cout R_{2}} &  - (\frac{1}{\cout R_2} + \frac{1}{\cout 
                               R_3}) \\
     \end{array} \right)
  \cdot
  \left(
  \begin{array}{c}
    \vint{} \\
    \vout \\
  \end{array} \right)
\end{align*}
which has the general solution
\begin{align*}
  \left(
  \begin{array}{c}
    \vint{} \\
    \vout \\
  \end{array} \right)
  =& c_1 \cdot
     \left(
     \begin{array}{c}
       \frac{1}{\cint{} R_2} \\
       \alpha + \beta \\
     \end{array} \right) e^{\lambda_1 t} +
  c_2 \cdot
  \left(
  \begin{array}{c}
    \frac{1}{\cint{} R_2} \\
    \alpha - \beta \\
  \end{array} \right) e^{\lambda_2 t},
\end{align*}
where
{\small
  \begin{flalign}
    &\alpha = \frac{\cout R_3 - \cint{}(R_2 + R_3)}{2 \cout \cint{} R_2 
      R_3},& \label{par1} \\
    &\beta=\frac{\sqrt{(\cout R_3 + \cint{} (R_2+ R_3))^2 - 4 \cout \cint{} 
        R_2 R_3}}{2 \cout \cint{} R_2 R_3},& \label{par2}\\
    &\lambda_{1,2}= - \frac{\cout R_3 + \cint{} (R_2+ R_3)}{2 \cout \cint{} 
      R_2 R_3} \pm \beta. & \label{par3}
  \end{flalign}
}	

\subsection{System $(0,1)$: $\va=\gnd$, $\vb=\vdd$}

Opening transistors $T_2$ and $T_3$, as shown in \cref{fig:nor_RC_case01},
decouples the nodes $N$ and $O$ once again. We thus get
the non-homogeneous system of ODEs

{\small
  \begin{align*}
    \cint{} \cdot {\operatorname{d}\over\operatorname{d}\!t} \vint{} &= 
                                                                       \iintm{} =
                                                                       \frac{\vdd-\vint{}}{R_1}, \\
    \cout \cdot {\operatorname{d}\over\operatorname{d}\!t} \vout &= \iout 
                                                                   =
                                                                   - \frac{\vout}{R_4},
  \end{align*}
}
which is in matrix representation
\begin{align*}
  \left(
  \begin{array}{c}
    {\operatorname{d}\over\operatorname{d}\!t} \vint{} \\
    {\operatorname{d}\over\operatorname{d}\!t} \vout \\
  \end{array} \right)
  =&
     \left(
     \begin{array}{cc}
       -\frac{1}{\cint{} R_{1}} & 0 \\
       0 &  - \frac{1}{\cout R_4} \\
     \end{array} \right)
  \cdot 
  \left(
  \begin{array}{c}
    \vint{} \\
    \vout \\
  \end{array} \right)\\
  +&
     \left(
     \begin{array}{c}
       \frac{\vdd}{\cint{} R_1} \\
       0 \\
     \end{array} \right).
\end{align*}

It is easy to check that the fundamental matrix solution to the 
corresponding homogeneous system is
\begin{align*}
  \phi(t)=
  \left(
  \begin{array}{cc}
    e^{-\frac{t}{\cint{} R_1}} & 0\\
    0 & e^{-\frac{t}{\cout R_4}} \\
  \end{array} \right)
\end{align*}
leading to the general non-homogeneous solution
\begin{align*}
  \left(
  \begin{array}{c}
    \vint{} \\
    \vout \\
  \end{array} \right)
  =&
     \left(
     \begin{array}{c}
       c_1 \cdot e^{-\frac{t}{\cint{} R_1}} + \vdd \\
       c_2 \cdot e^{-\frac{t}{\cout R_4}} \\
     \end{array} \right).
\end{align*}

\subsection{System $(0,0)$: $\va=\gnd$, $\vb=\gnd$}

Closing both \pmos\ transistors, as shown in \cref{fig:nor_RC_case00},
causes both capacitances to be charged over the same resistor $R_1$,
similarly to system (1,0). Since $\iout=I_1-\iintm{}$, the
ODE system describing the behavior is
{\small
  \begin{align*}
    \cout \cdot {\operatorname{d}\over\operatorname{d}\!t} \vout &= \iout =
                                                                   \frac{\vint{} - \vout}{R_2},\\
    \cint{} \cdot {\operatorname{d}\over\operatorname{d}\!t} \vint{} &= 
                                                                       \iintm{} =
                                                                       I_1 - \iout
                                                                       = \frac{\vdd-\vint{}}{R_1} - \frac{\vint{} - \vout}{R_2},
  \end{align*}
}
which is a non-homogeneous system, with the matrix form
\begin{align*}
  \left(
  \begin{array}{c}
    {\operatorname{d}\over\operatorname{d}\!t} \vint{} \\
    {\operatorname{d}\over\operatorname{d}\!t} \vout \\
  \end{array} \right)
  =&
     \left(
     \begin{array}{cc}
       -(\frac{1}{\cint{} R_1} + \frac{1}{\cint{} R_2}) & \frac{1}{\cint{} 
                                                          R_2} \\
       \frac{1}{\cout R_2} &  - \frac{1}{\cout R_2} \\
     \end{array} \right)
  \cdot \\
   &\left(
     \begin{array}{c}
       \vint{} \\
       \vout \\
     \end{array} \right)
  +
  \left(
  \begin{array}{c}
    \frac{\vdd}{\cint{} R_1} \\
    0 \\
  \end{array} \right).
\end{align*}
By straightforward but tedious calculations, it follows that the 
fundamental matrix solution of the homogeneous system is
\begin{align*}
  \phi(t)=
  \left(
  \begin{array}{cc}
    \frac{1}{\cint{} R_2} \cdot e^{\lambda_1 t} & \frac{1}{\cint{} R_2} 
                                                  \cdot e^{\lambda_2 t}\\
    (\alpha + \beta) \cdot e^{\lambda_1 t} & (\alpha - \beta) \cdot 
                                             e^{\lambda_2 t} \\
  \end{array} \right),
\end{align*}
where
{\small
  \begin{flalign}
    &\alpha = \frac{\cout(R_1 + R_2) - \cint{} R_1 }{2 \cout \cint{} R_1 
      R_2},& \label{par4}\\
    &\beta=\frac{\sqrt{(\cint{} R_1 + \cout (R_1+ R_2))^2 - 4 \cout \cint{} 
        R_1 R_2}}{2 \cout \cint{} R_1 R_2},& \label{par5}\\
    &\gamma= - \frac{ \cint{} R_1 + \cout (R_1 + R_2)}{2 \cout \cint{} R_1 
      R_2},& \label{par6}\\
    &\lambda_{1,2}= \gamma \pm \beta. \label{par7}&
  \end{flalign}
}	

\begin{figure}[t]
  \centering
  \includegraphics[width=0.8\linewidth]{\figPath{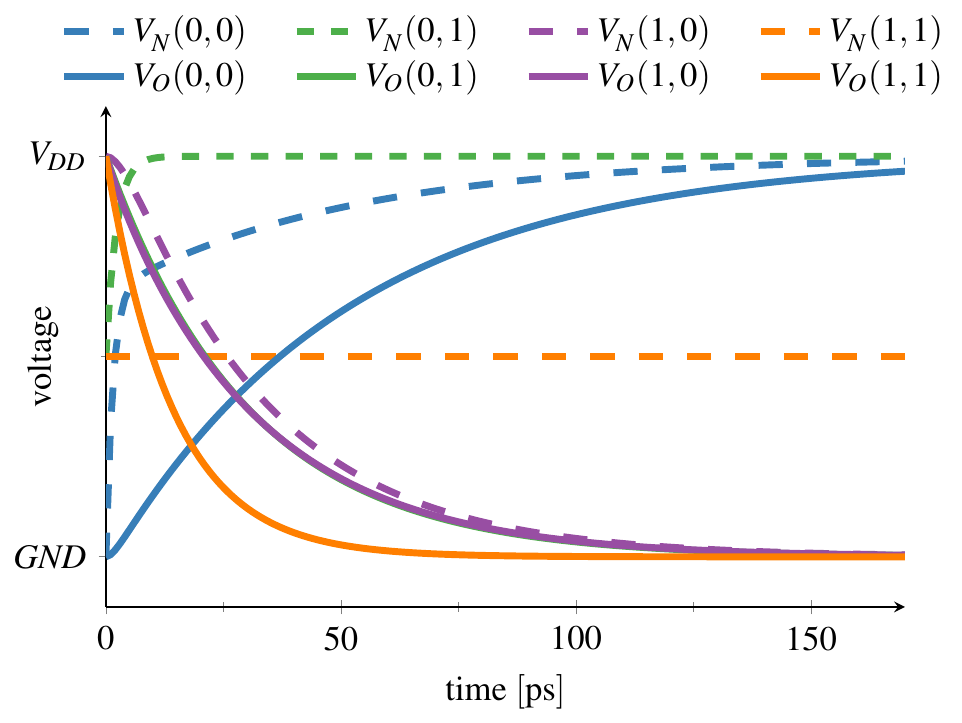}}
  \caption{Temporal evolution of the trajectories for all systems.}
  \label{fig:systems_plot}
\end{figure}	

The general solution of the non-homogeneous system is
\begin{align*}
  \left(
  \begin{array}{c}
    \vint{} \\
    \vout \\
  \end{array} \right)
  =&
     \left(
     \begin{array}{c}
       \frac{c_1}{\cint{} R_2} e^{\lambda_1 t} + \frac{c_2}{\cint{} R_2} 
       e^{\lambda_2 t} +\vdd \\
       c_1 \cdot (\alpha + \beta) e^{\lambda_1 t} + c_2 \cdot (\alpha - 
       \beta) e^{\lambda_2 t} + \vdd \\
     \end{array} \right). \\
\end{align*}

\subsection{Trajectory Comparison}

\cref{fig:systems_plot} depicts the signals $V_{N/O}(n,m)(t)$ over time $t$ in
system $(n,m)$. The initial values were set to $\vint{}(0)=\vout(0)=\vdd$, with
the exception of $\vint{}(0,0)(0)=\vout(0,0)(0)=\gnd$ and
$\vint{}(1,1)(0)=\vdd/2$. Compared to the cases where only one \nmos\ is closed,
the output trajectory of system $(1,1)$ is much steeper. Note that this
is in line with the considerations for the speed-up MIS effect in
\cref{sec:SPIC}.

\section{Modeling MIS Effects}
\label{sec:charlie}

In this section, we will investigate how well our simple hybrid ODE model is
capable of faithfully representing the MIS effects described in \cref{sec:SPIC}.
It turns out that the speed-up is modeled appropriately, whereas the
slow-down is only partially covered. Note that also the approaches presented
in~\cite{Melcher92:MAM, AB06:SST} struggled with this effect, such that the
authors finally resorted to fitting the delays for these cases.

For our analysis, we computed the delay as a function of the input separation
time $\Delta=t_B-t_A$ for falling and rising output transitions, and compared it
with our analog simulation results (cf. \cref{fig:nor2_out_down_charlie} and
\cref{fig:nor2_out_up_charlie}). Similar to \cref{sec:SPIC}, we start with a
falling output transition.  To compute the delay for a given $\Delta$, we need
to combine two solutions:
\paragraph*{1)}Starting in the system $(0,0)$ initially, which models a gate
whose inputs have been $0$ for a very long time, we switch to $(1,0)$ resp.\
$(0,1)$ at time $t=0$ and compute the corresponding trajectory.
\paragraph*{2)}When in the mode entered in $1)$, we switch to system $(1,1)$ at
time $t_s$, and determine $t_O$ where $\vout(t_O)=\vdd/2$. The delay is
extracted as $t_O - \min(t_A,t_B)= t_O$, since the earlier of the two inputs
triggers the output transition. Starting in system $(1,0)$ results in
$\Delta=t_s$, whereat for system $(0,1)$ we get $\Delta=-t_s$, which accounts
for the reversed order of the input transitions.

The calculation of the rising output delay is carried out analogously, with the
exception that we start in the system $(1,1)$ initially and switch to $(1,0)$
($\Delta <0$) resp.\ $(0,1)$ ($\Delta > 0$) at $t=0$, before turning to $(0,0)$
at $t_s$. The sought delay is now equal to $t_O - \max(t_A,t_B)=t_O-t_s$, since
it is the later of the two inputs that initiates the output change. Note
carefully, however, that it is unfortunately not clear which initial value to
use for $\vint{}$ in the system $(1,1)$ here:  As the latter does not change the
value of $\vint{}(t)$ at all, the proper initial value would be the
\emph{actual} value of $\vint{}$ in the state $(m,n)$ the system was in
\emph{before/at} the switch to $(1,1)$ occurred.

For a quantitative comparison, we parameterized the resistances and capacitances
in our model using a least square fitting approach, with the goal to match the
output threshold crossing times $\dsu(\pm\infty)$ and $\dsu(0)$
resp. $\dsd(\pm\infty)$ and $\dsd(0)$ shown in \cref{fig:nor_analog_sim}.
Interestingly, simultaneous fitting of all three data points turned out to be
impossible, even for the ``well-behaved'' case of falling output transitions.
To understand why, we derived analytic expressions for the values
\[
\ddoD(-\infty)\approx \ln(2)\cdot \cout R_4 \quad\mbox{and}\quad
\ddoD(0)=\frac{\ln(2) \cdot\cout R_3R_4}{R_3 + R_4}
\]
(as well as for $\ddoD(\infty)$, $\dupD(-\infty)$, $\dupD(0)$ and
$\dupD(\infty)$), by inverting the explicit formulas of our
trajectories. Unfortunately, necessary simplifications that enabled these
calculations induced some approximation errors, which made a direct computation
of the desired parameters impossible.

\begin{table}[t]
\centering
\caption{Empirically obtained parameter values}
\label{Table:Param}
  \renewcommand{\arraystretch}{1.2}
\sisetup{
  round-mode = places,
  round-precision =3,
  scientific-notation = engineering,
  table-figures-decimal=3,
  table-figures-integer=2,
}
\begin{tabular}{S S}
{Parameter} & {Value}                                \\
\hline
{$R_1$}     & \SI{37088.32043327145}{\ohm}             \\
{$R_2$}     & \SI{44925.83293787842}{\ohm}           \\
{$R_3$}     & \SI{45149.85667051946}{\ohm}            \\
{$R_4$}     & \SI{48761.4927022873}{\ohm}            \\
{$\cint{}$}     & \SI{5.948581669628511e-17}{\farad} \\
{$\cout$}     & \SI{6.172588967251559e-16}{\farad}%
\end{tabular}
\end{table}

Nevertheless, important insights could be gained. More specifically, since $R_3$
and $R_4$ are the on-resistors of the two nMOS transistors and should hence be
roughly the same, we obtain
$\frac{\ddoD(-\infty)}{\ddoD(0)} \approx \frac{R_3+R_4}{R_3} \approx 2$.  Since
the desired ratio according to our simulations (see \cref{fig:nor_analog_sim})
is $\frac{\dsd(-\infty)}{\dsd(0)} \approx \frac{\SI{38}{\ps}}{\SI{28}{\ps}}$,
however, we could not even simultaneously fit these two values with reasonable
choices for $R_3$ and $R_4$.

We solved the problem by adding (that is, subtracting) a pure delay 
$\dmin=\SI{18}{\ps}$, also present in the original IDM,
which defers the switching to the new state upon an input transition.
This results in an effective ratio of $\frac{\SI{20}{\ps}}{\SI{10}{\ps}}=2$,
which could finally be matched by least squares fitting, leading to the 
parameter values presented in \cref{Table:Param}. To also 
accommodate $\dmin$ in the MIS delay computations, we just need to transform
$t_O - \min(t_A,t_B)= t_O$ to $\dmd(\Delta)=t_O+\dmin$ and, correspondingly,
$\dmu(\Delta)=t_O - t_s + \dmin$. Note that
the same $\dmin=\SI{18}{\ps}$ was also used for rising output transitions.

\begin{figure}[tbp]
  \centering
  \includegraphics[width=0.75\linewidth]{\figPath{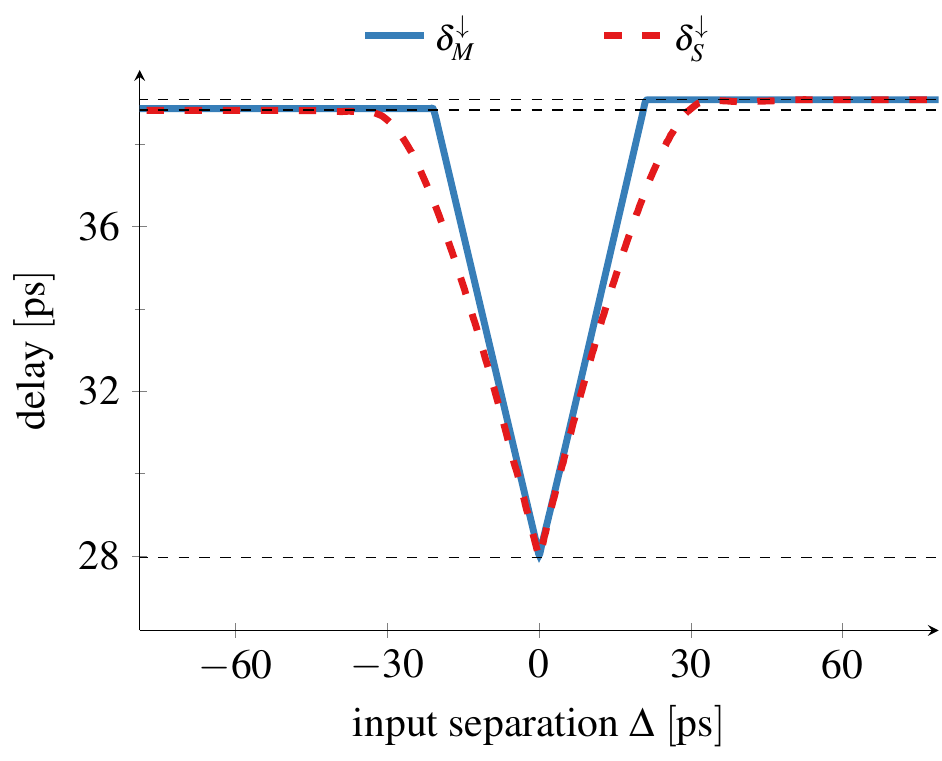}}
  \caption{Computed MIS delays for falling output transitions.}
  \label{corFig3}
\end{figure}

Utilizing the found parameters, we can finally visualize the delay predictions
of our model. \cref{corFig3} shows the very good fit of $\dmd(\Delta)$ for a falling
output transition compared to the analog simulation results presented in
\cref{sec:SPIC}.  Unfortunately, a comparable coverage of the MIS effects for
rising output transitions cannot be achieved (see \cref{corFig5}): 
For none of the initial values $\vint{} \in \{\gnd,\vdd/2,\vdd\}$, the computed
delay $\dmu(\Delta)=t_O - t_s + \dmin$ reasonably matches
$\dsu(\Delta)$ obtained in our analog simulations. More specifically, for
$\vint{}=\vdd$ and $\vint{}=\vdd /2$, our simple ODE model fails to correctly
predict the case of $\Delta<0$, i.e., switching to system $(1,0)$ at $t=0$ and
then to $(0,0)$ at time $t=\Delta$. In the case of $\vint{}=\gnd$, which
reasonably matches $\dsu(\pm\infty)$ and $\dsd(\pm\infty)$ and thus has been
used in \cref{sec:modelingaccuracy}, it fails to model the MIS peak around
$\Delta=0$ in \cref{fig:nor2_out_up_charlie} for small negative and positive
$\Delta$.

We conclude that our simple ODE model perfectly captures the MIS effects
caused by the parallel transistors, but not for the ones caused by transitors
arranged in series.

\section{Parametrization}
\label{sec:param}

\begin{figure}[tbp]
  \centering
  \includegraphics[width=0.9\linewidth]{\figPath{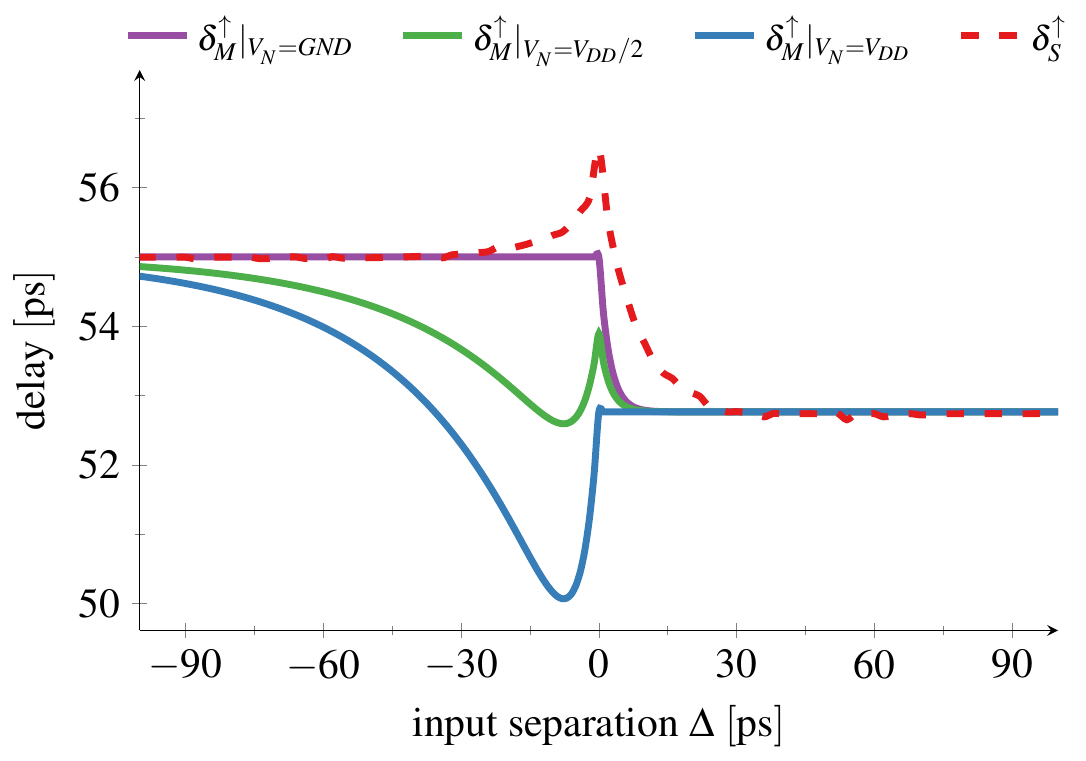}}
  \caption{Computed MIS delays for rising output transitions.}
\label{corFig5}
\end{figure}

In order to apply our hybrid model in practice, in particular, for 
digital timing simulations using the Involution Tool, one needs
to compute the input-to-output delay functions for all possible
state transitions, including the ones shown in 
\cref{corFig3} and \cref{corFig5}.
This, in turn, requires a proper \emph{parametrization} of our model,
i.e., the determination of the parameters $R_1,\dots,R_4,\cint,\cout$
that cause our model to match the actual delays of a given NOR gate 
in a circuit as good as possible.

There is no unique and possibly even optimal parametrization 
approach, but it is natural to match \cref{corFig3} and \cref{corFig5}
(for $\Delta>0$) to \cref{fig:nor2_out_down_charlie} and 
\cref{fig:nor2_out_up_charlie} as close as possible. For this purpose, it 
is sufficient to
match the \emph{characteristic} Charlie delay values $\ddoD(-\infty)$, 
$\ddoD(0)$, $\ddoD(\infty)$ in \cref{fig:nor2_out_down_charlie} 
(falling output transition) and the corresponding values $\dupD(-\infty)$, 
$\dupD(0)$, $\dupD(\infty)$ in \cref{fig:nor2_out_up_charlie}.

Below, we provide exact or approximate\footnote{Note that all the errors 
  related to the approximations we describe in this section is in $O(t^2)$, 
  where $t$ is very small ($t\leq 2 \times 10^{-10}$). The error is hence so small 
  that it can be ignored in practice.} analytic formulas\footnote{It is worth noting
  that in parallel with obtaining these equations, we used built-in MATLAB software functions 
  (e.g. the non-linear optimization function \emph{fminbnd}) to validate the equations.}
for the characteristic Charlie delay values in 
\cref{corFig3} and \cref{corFig5} in terms of the 
parameters ($R_1$-$R_4$, $\cint{}$, and $\cout$). These 
expressions will allow us to understand which parameters 
affect the which value, and could even be used for explicit parametrization
of a circuit with given characteristic Charlie delays.

For falling output transitions (\cref{corFig3}, $\Delta=0$), 
an exact formula for $\ddoD(0)$ is
\begin{equation}
  \ddoD(0)=\frac{-\ln(0.5)}{\frac{1}{\cout R_3}+\frac{1}{\cout 
      R_4}}\label{eq:ddoD0}
\end{equation}

For falling output transitions (\cref{corFig3}, $\Delta=-\infty$)
an exact formula for computing $\ddoD(-2 \times 10^{-10})\approx 
\ddoD(-\infty)$ is
\begin{equation}
  \ddoD(-2 \times 10^{-10})=-\ln(0.5)\cdot \cout 
  R_4\label{eq:ddoDminus}
\end{equation}

For falling output transitions (\cref{corFig3}, $\Delta=\infty$), 
an approximation to compute $d=\ddoD(2 \times 
10^{-10})\approx\ddoD(\infty)$ 
is
\begin{equation}
  d \approx \frac{0.6-[c_1(\alpha+\beta)e^{\lambda_1 w}(1- \lambda_1 
    w)+c_2(\alpha-\beta)e^{\lambda_2 w}(1-\lambda_2 
    w)]}{c_1(\alpha+\beta)\lambda_1 e^{\lambda_1 
      w}+c_2(\alpha-\beta)\lambda_2e^{\lambda_2 w}},\label{eq:ddoDplus}
\end{equation}
where $\alpha$, $\beta$, $\lambda_{1,2}$ are given in (\ref{par1}), 
(\ref{par2}), and (\ref{par3}), respectively, and 
\begin{flalign*}
  &w=10^{-10},&\\
  &c_2=\frac{0.6[((\alpha+\beta)\cint{} R_2)-1]}{\beta},&\\
  &c_1= (\vdd \cint{} R_2)-c_2. &
\end{flalign*}

For rising output transitions (\cref{corFig5}, 
$\dupD(\Delta)$ for any $\Delta \geq0$ and initial value 
$\vint(0)=X$), 
an approximation for $d=\dupD(\Delta)$ is
\begin{align}
  d &\approx \frac{0.6 - l -c_1^{\Delta} \cdot (\alpha+\beta) e^{\lambda_1 
      w} (1 - \lambda_1 w)}{c_1^{\Delta} \cdot 
      (\alpha+\beta) \lambda_1 e^{\lambda_1 w} + c_2^{\Delta} \cdot 
      (\alpha - \beta) \lambda_2 e^{\lambda_2 w}} \label{eq:dupDplus}\\
    &\quad -
      \frac{c_2^{\Delta} \cdot (\alpha - \beta) 
      e^{\lambda_2 w} (1 - \lambda_2 w)}{c_1^{\Delta} \cdot 
      (\alpha+\beta) \lambda_1 e^{\lambda_1 w} + c_2^{\Delta} \cdot 
      (\alpha - \beta) \lambda_2 e^{\lambda_2 w}} - 
      \Delta,\nonumber
\end{align}
where $\alpha$, $\beta$, $\gamma$, and $\lambda_{1,2}$ are respectively 
equal to (\ref{par4}), (\ref{par5}), (\ref{par6}), and (\ref{par7}). 
Moreover, we have
\begin{flalign*}
  &w=2\times 10^{-10},&\\
  &l=\frac{\vdd(-\alpha^2+\beta^2)R_2}{R_1(\gamma^2- \beta^2)},&\\
  &c_2^{\Delta} = \frac{([(\alpha+\beta) 
    \vint{{(0,1)}}(\Delta)]+a+b)\cint{} R_2}{2 \beta e^{\lambda_2 
      \Delta}},&\\
  &c_1^{\Delta} = \frac{[(\alpha+\beta) \vint{{(0,1)}}(\Delta) - c_2 
    \frac{\alpha + \beta}{\cint{} R_2} e^{\lambda_2 \Delta} +a] \cint{} 
    R_2}{(\alpha + \beta) e^{\lambda_1 \Delta}},&\\
  &a=\frac{\vdd(\alpha+ \gamma)(\alpha+\beta)}{\cint{} R_1(\gamma^2- 
    \beta^2)},&\\
  &b= \frac{\vdd(-\alpha^2 + \beta^2)}{\cint{} R_1(\gamma^2- 
    \beta^2)},&\\
  &\vint{{(0,1)}}(\Delta)=\vdd + (X-\vdd) e^{\frac{-\Delta}{\cint{} 
      R_1}}.&
\end{flalign*}
Recall that one does not usually have information about the initial 
value $\vint{{(0,1)}}(0)=X$; \cref{corFig5} shows $X=0$, $X=\vdd/2$ and $X=\vdd$.

For rising output transitions (\cref{corFig5}, 
$\dupD(\Delta)$ for any $\Delta < 0$ and initial value $\vint(0)=X$), 
an approximation for $d=\dupD(\Delta)$ is
\begin{align}
  d &\approx \frac{0.6-l - c_1^{\Delta} \cdot (\alpha+\beta) e^{\lambda_1 
      w} (1 - \lambda_1 w)}{c_1^{\Delta} \cdot 
      (\alpha+\beta) \lambda_1 e^{\lambda_1 w} + c_2^{\Delta} \cdot 
      (\alpha - \beta) \lambda_2 e^{\lambda_2 w}}\label{eq:dupDminus}\\ 
    & \quad - \frac{c_2^{\Delta} \cdot (\alpha - \beta) 
      e^{\lambda_2 w} (1 - \lambda_2 w)}{c_1^{\Delta} \cdot 
      (\alpha+\beta) \lambda_1 e^{\lambda_1 w} + c_2^{\Delta} \cdot 
      (\alpha - \beta) \lambda_2 e^{\lambda_2 w}} - 
      |\Delta|,\nonumber
\end{align}
where
$l$, $\alpha$, $\beta$, $\gamma$, $\lambda_{1,2}$, $a$, and $b$, are as 
same as Case 4 and
\begin{flalign*}
  &w=10^{-10},&\\
  &c_2^{\Delta} = \frac{([(\alpha+\beta) \vint{{(1,0)}}(\Delta)- 
    \frac{\vout^{(1,0)}(\Delta)}{\cint{} R_2}]+a+b)\cint{} R_2}{2 \beta 
    e^{\lambda_2 \Delta}},&\\
  &c_1^{\Delta} = \frac{[(\alpha+\beta) \vint{{(1,0)}}(\Delta) - c_2 
    \frac{\alpha + \beta}{\cint{}R_2} e^{\lambda_2 \Delta} +a]\cint{} 
    R_2}{(\alpha + \beta) e^{\lambda_1 \Delta}},&\\
  &\vint{(1,0)}(\Delta)=\frac{g_1}{\cint{} R_2}e^{(z+y)\Delta}+ 
  \frac{g_2}{\cint{} R_2} e^{(z-y) \Delta},&\\
  &\vout^{(1,0)}(\Delta)=g_1(x+y)e^{(z+y)\Delta}+ g_2(x-y)e^{(z-y) 
    \Delta},&\\
  &z= - \frac{\cout R_3+D(R_2+R_3)}{2\cout \cint{} R_2R_3},&
\end{flalign*}
with $x=\alpha$ and $y=\beta$ given by (\ref{par1}) and (\ref{par2})
and $g_1=\frac{(y-x)g_2}{x+y}$.

Again, since the initial value of $X=\vint{{(1,0)}}(0)$ is usually 
unknown, 
we consider only the cases $X=0$, $X=\vdd/2$ and $X=\vdd$ shown in
\cref{corFig5}, which lead to
\begin{itemize}
\item if $X=0$, then $g_2=0$,
\item If $X=\vdd$, then $g_2=\frac{0.6(x+y)DR_2}{y}$,
\item If $X=\vdd/2$, then $g_2=\frac{0.3(x+y)DR_2}{y}$.
\end{itemize}

It is apparent from \eqref{eq:ddoD0}--\eqref{eq:ddoDplus} that the
characteristic Charlie delays in \cref{corFig3} are not affected by $R_1$ at
all. To be more precise, $\ddoD(0)$ is determined by $\cout$, $R_3$, and
$R_4$, while $\ddoD(-\infty)$ is determined by $\cout$ and $R_4$ only;
$\ddoD(\infty)$ is affected by $\cint{}$, $\cout$, $R_2$, and $R_3$. On the
other hand, the characteristic Charlie delays $\dupD(0)$ and $\dupD(\infty)$
in \cref{corFig5}, which seems to match \cref{fig:nor2_out_up_charlie} best,
are only affected by $\cint{}$, $\cout$, $R_1$, and $R_2$ according to
\cref{eq:dupDplus}.  Finally, \cref{eq:dupDminus} reveals that
$\dupD(-\infty)$ does not depend on $R_4$.

One immediate consequence of this is that $R_3$ and $R_4$ are fixed already by
matching $\ddoD(0)$ and $\ddoD(-\infty)$. To simultaneously match
$\ddoD(\infty)$, in theory, $\cint{}$ and $R_2$ were still available, but
since their influence on $\ddoD(\infty)$ is very small, this is not always
effective.  Even worse, according to \cref{eq:ddoD0} and \cref{eq:ddoDminus},
it may even be impossible to find a parametrization for $R_3$ and $R_4$ that
allows to match even the two characteristic Charlie delay values $\ddoD(0)$
and $\ddoD(-\infty)$ simultaneously, which happens in the case of too large a
ratio of $\ddoD(-\infty)/\ddoD(0)$.  And indeed, as already mentioned, a pure
delay of $\dmin=18$~ps had to be subtracted from all these delay values in
order to be able to determine a matching set of parameters.

We conclude this section by noting that relying on the characteristic Charlie
delays from \cref{corFig5} for $\vint\ = X=0$ for parametrization makes sense,
since the system $(0,0)$ starts, at time $\Delta$, from $V_{DD}$ in this case,
i.e., a value not affected by $X$.  This is in accordance with the fact that
our model does reasonably capture the Charlie-effect in
\cref{fig:nor2_out_up_charlie} for the case $\Delta \geq 0$ (that is, for
$X=0$). It is apparent, though, that the parasitic capacitance $\cint{}$ has a
substantial influence on the characteristic Charlie delays, as it is a factor
in the denominator of $\dupD(0)$ and $\dupD(\infty)$. Like the invariance of
$\vint{}$ in the system $(1,1)$, which has already been identified in
\cref{sec:charlie} as the main cause for our model's inability to fully cover
the Charlie effect, this is an unwanted artefact of our simplistic modeling.
	
\section{Modeling Accuracy}
\label{sec:modelingaccuracy}

In this section, we will compare our hybrid model to inertial delays and the
IDM.  To be able to do so, we added our model to the publicly available
Involution Tool \cite{OMFS20:INTEGRATION}.  Previously, all channel
implementations had to be written in VHDL.  However, since this approach was
already rather tedious for a SumExp-Channel, where the inverse of the trajectory
had to be numerically approximated, we decided to come up with a new way of
implementing channels: Using the QuestaSim Foreign Language Interface (FLI)
\cite{MSIM:FLI} allowed us to escape the Involution Tool's standard VHDL
environment and to execute C code. In a second step, Python code was called from
C, which finally implemented our hybrid channels.  This approach enabled us to
utilize the complete Python ecosystem, which reduces the effort the hybrid
channel implementation itself drastically.
	
For the evaluation, we again used the \SI{15}{\nm} Nangate Open Cell Library
featuring FreePDK15$^\text{TM}$ FinFET models~\cite{Nangate15}
($\vdd=\SI{0.8}{\V}$).  Based on a Verilog description of a \NOR\ gate, we
utilized the Cadence tools Genus and Innovus (version 19.11) to perform
optimization, placement and routing.  Finally, we extracted the parasitic
networks from the final layout to obtain \spice\ models, which we used as golden
reference in analog Spectre (version 19.1) simulations.

\begin{figure}[t]
  \centering
  \includegraphics[width=1\linewidth]{\figPath{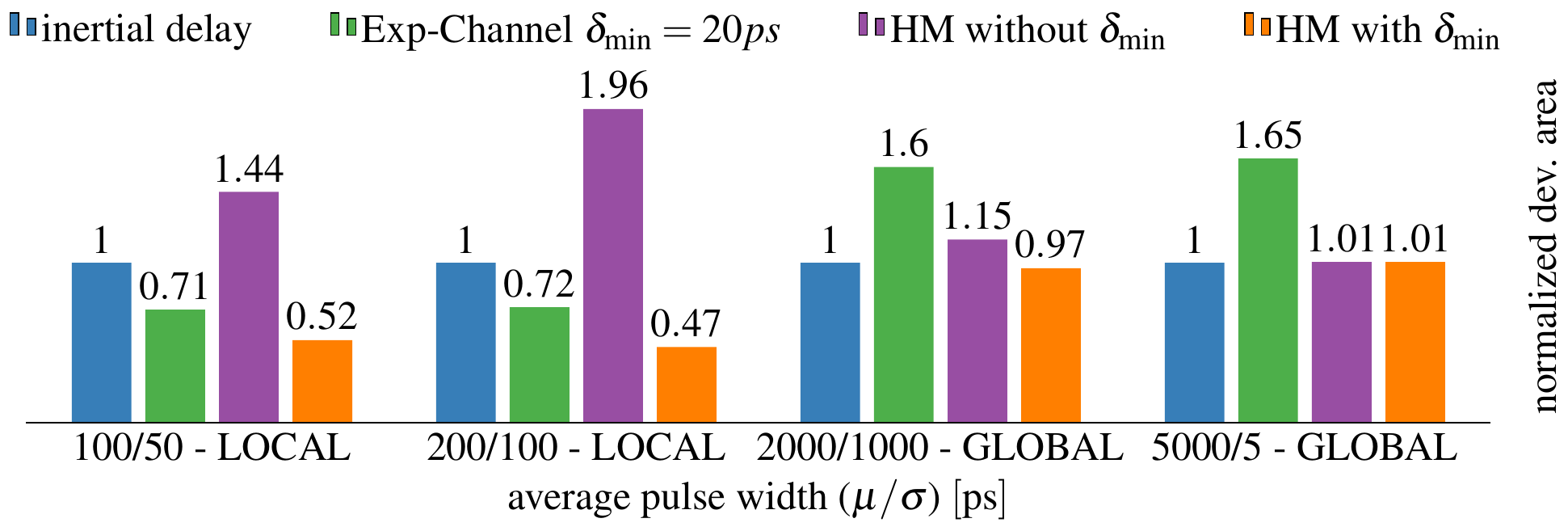}}
  \caption{\label{fig:nor_results} Accuracy of inertial delay,
    Exp-Channel and hybrid model compared to analog simulations of
    a \NOR\ gate. Lower bars indicate better results.}
\end{figure}

Using the parameter set introduced in \cref{Table:Param}, we performed
simulations for various waveform configurations, ranging from very short to
broad pulses. Each simulation consisted of 500 transitions, except for the last
simulation, where we generated 250 transitions.  The simulations have been
repeated 20 times, and the averaged results are presented in
\cref{fig:nor_results}.  The waveform configuration \emph{100/50 - LOCAL}
describes the case where transitions are created individually for each input,
according to a normal distribution, with $\mu = \SI{100}{\ps}$ and
$\sigma = \SI{50}{\ps}$.  \emph{GLOBAL} indicates that the transitions are not
calculated separately for each input but rather for all inputs together. This
option allows to test how accurately the delay models perform for large absolute
values of $\Delta$, since concurrent transitions are unlikely with this
configuration.
	
The results are compared in terms of the deviation area, which is calculated as
follows: The digitized \spice\ traces are subtracted from the corresponding
traces of the digital delay model and the absolute area is summed up. Since
absolute values are meaningless, the results are normalized with the inertial
delay as baseline.
	
\begin{figure}[t]
  \centering
  \includegraphics[width=0.75\linewidth]{\figPath{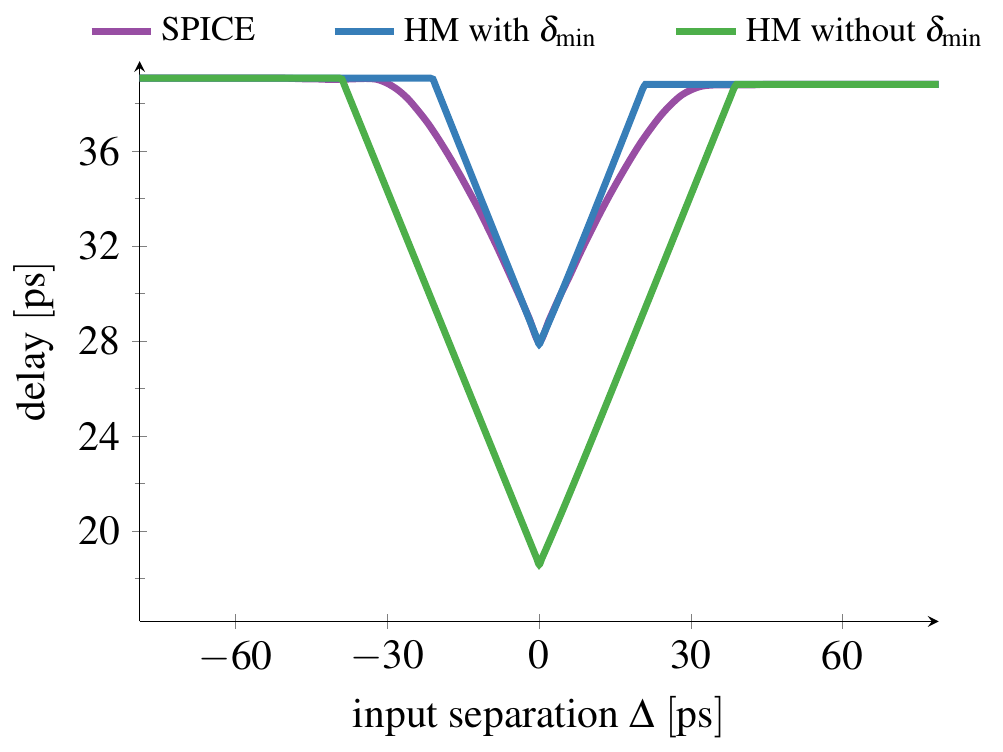}}
  \caption{Matching of the hybrid model with and without pure delay for 
    falling output transitions.}
  \label{fig:matching_falling_output}
\end{figure}

For short pulses ($\mu = \SI{100}{\ps}, \mu= \SI{200}{\ps}$), the superiority of
the hybrid model with $\dmin$ can be clearly seen.  The deviation area is less
than half that of the inertial delays.  Moreover, it also outperforms the
Exp-Channel, which we chose as representation for the IDM in the Involution
Tool, with $\dmin = \SI{20}{\ps}$.  Note that we had to determine the latter
empirically, since there is no proper parametrization of IDM channels
representing multi-input gates available.  The hybrid model without pure delay
performs worse, which is primarily due to the imperfect delay matching, which
can be seen in \cref{fig:matching_falling_output}. Note that the pure delay
shows no effect for rising input transitions. We should also emphasize that,
for the first two waveform configurations, a lot of transitions are happening
within a range of $\Delta = [\SI{-40}{\ps}, \SI{40}{\ps}]$, where the hybrid
model without pure delay has deficiencies.

For broader pulses ($\mu = \SI{2}{\ns}, \mu= \SI{5}{\ns}$), which are covered by
the last two waveform configurations, it can be seen that the hybrid model and
the inertial delay model perform similar.  This is due to the fact that
$|\Delta| \gg \SI{100}{\ps}$, where the matching is nearly perfect.  The
Exp-Channel shows deficiencies for broad pulses, which is caused by placing the
delay channel at the output and the consequential inability to determine which
input caused the transition. Since $\dsu(\infty)$ and $\dsu(-\infty)$ differ,
(cf. \cref{corFig3} and \cref{corFig5}) the Exp-Channel simply delays the
transition by their average, which explains the observed inaccuracies.

In terms of simulation runtime, our simple experiments reveal a minor overhead
of the hybrid model compared to the simple inertial delay model or the
Exp-Channel of \SI{6}{\percent}, which seems acceptable in view of the increased
modeling accuracy. For more robust numbers, more extensive simulation runs are
necessary, which we are planning to execute in the near future.
	
\section{Conclusions}
\label{sec:conclusions}
	
We introduced a simple hybrid ODE model for a two-input NOR gate, which
naturally generalizes the hybrid analog model corresponding to standard
single-input, single-output involution channels. The ODEs governing the
switching waveforms of the output, based on the state of the inputs, have been
obtained by replacing transistors with ideal switches in a simple RC model of
the circuit. By analytically solving the resulting ODE systems, we obtained a
digital gate delay model that faithfully reproduces all MIS effects, except in
one particular situation.  We also incorporated our hybrid model in the
Involution Tool for digital timing analysis and compared the average accuracy
for random traces in a custom circuit for different channel models. Our results
show that our new hybrid model outperforms both classic involution channels and
standard inertial delay channels with respect to modeling accuracy.

Future work will be devoted to the question of whether our multi-input digital
delay channels are continuous with respect a certain metric, and therefore lead
to a faithful model. In addition, we will look out for alternative models that
fully capture all MIS effects.



\begin{thebibliography}{10}
\providecommand{\url}[1]{#1}
\csname url@samestyle\endcsname
\providecommand{\newblock}{\relax}
\providecommand{\bibinfo}[2]{#2}
\providecommand{\BIBentrySTDinterwordspacing}{\spaceskip=0pt\relax}
\providecommand{\BIBentryALTinterwordstretchfactor}{4}
\providecommand{\BIBentryALTinterwordspacing}{\spaceskip=\fontdimen2\font plus
\BIBentryALTinterwordstretchfactor\fontdimen3\font minus
  \fontdimen4\font\relax}
\providecommand{\BIBforeignlanguage}[2]{{%
\expandafter\ifx\csname l@#1\endcsname\relax
\typeout{** WARNING: IEEEtran.bst: No hyphenation pattern has been}%
\typeout{** loaded for the language `#1'. Using the pattern for}%
\typeout{** the default language instead.}%
\else
\language=\csname l@#1\endcsname
\fi
#2}}
\providecommand{\BIBdecl}{\relax}
\BIBdecl

\bibitem{Ung71}
S.~H. Unger, ``Asynchronous sequential switching circuits with unrestricted
  input changes,'' \emph{IEEE Transaction on Computers}, vol.~20, no.~12, pp.
  1437--1444, 1971.

\bibitem{BJV06}
M.~J. Bellido-D{\'{\i}}az, J.~Juan-Chico, and M.~Valencia, \emph{Logic-Timing
  Simulation and the Degradation Delay Model}.\hskip 1em plus 0.5em minus
  0.4em\relax London: Imperial College Press, 2006.

\bibitem{FNNS19:TCAD}
M.~{Függer}, R.~{Najvirt}, T.~{Nowak}, and U.~{Schmid}, ``A faithful binary
  circuit model,'' \emph{IEEE Transactions on Computer-Aided Design of
  Integrated Circuits and Systems}, vol.~39, no.~10, pp. 2784--2797, 2020.

\bibitem{FNS16:ToC}
M.~F\"ugger, T.~Nowak, and U.~Schmid, ``Unfaithful glitch propagation in
  existing binary circuit models,'' \emph{IEEE Transactions on Computers},
  vol.~65, no.~3, pp. 964--978, March 2016.

\bibitem{OMFS20:INTEGRATION}
D.~Öhlinger, J.~Maier, M.~Függer, and U.~Schmid, ``The involution tool for
  accurate digital timing and power analysis,'' \emph{Integration}, vol.~76,
  pp. 87 -- 98, 2021.

\bibitem{CGB01:DAC}
L.-C. Chen, S.~K. Gupta, and M.~A. Breuer, ``A new gate delay model for
  simultaneous switching and its applications,'' in \emph{Proceedings of the
  38th Design Automation Conference (IEEE Cat. No.01CH37232)}, 2001, pp.
  289--294.

\bibitem{SRC15:TDAE}
\BIBentryALTinterwordspacing
A.~R. Subramaniam, J.~Roveda, and Y.~Cao, ``A finite-point method for efficient
  gate characterization under multiple input switching,'' \emph{ACM Trans. Des.
  Autom. Electron. Syst.}, vol.~21, no.~1, pp. 10:1--10:25, Dec. 2015.
  [Online]. Available: \url{http://doi.acm.org/10.1145/2778970}
\BIBentrySTDinterwordspacing

\bibitem{SKJPC09:ISOCC}
J.~Shin, J.~Kim, N.~Jang, E.~Park, and Y.~Choi, ``A gate delay model
  considering temporal proximity of multiple input switching,'' in \emph{2009
  International SoC Design Conference (ISOCC)}, Nov 2009, pp. 577--580.

\bibitem{SC06:DATE}
J.~Sridharan and T.~Chen, ``Modeling multiple input switching of cmos gates in
  dsm technology using hdmr,'' in \emph{Proceedings of the Design Automation
  Test in Europe Conference}, vol.~1, March 2006, pp. 6 pp.--.

\bibitem{RS21:TCAD}
O.~V.~S. Shashank~Ram and S.~Saurabh, ``Modeling multiple-input switching in
  timing analysis using machine learning,'' \emph{IEEE Transactions on
  Computer-Aided Design of Integrated Circuits and Systems}, vol.~40, no.~4,
  pp. 723--734, 2021.

\bibitem{Melcher92:MAM}
\BIBentryALTinterwordspacing
E.~Melcher, W.~Röthig, and M.~Dana, ``Multiple input transitions in cmos
  gates,'' \emph{Microprocessing and Microprogramming}, vol.~35, no.~1, pp. 683
  -- 690, 1992, software and Hardware: Specification and Design. [Online].
  Available:
  \url{http://www.sciencedirect.com/science/article/pii/016560749290387M}
\BIBentrySTDinterwordspacing

\bibitem{AB06:SST}
N.~Abdallah and P.~Bazargan-Sabet, ``Modeling the effects of input slew rate
  and temporal proximity of input transitions in event-driven simulation,'' in
  \emph{2006 Proceeding of the Thirty-Eighth Southeastern Symposium on System
  Theory}, March 2006, pp. 185--189.

\bibitem{Nangate15}
\BIBentryALTinterwordspacing
M.~Martins, J.~M. Matos, R.~P. Ribas, A.~Reis, G.~Schlinker, L.~Rech, and
  J.~Michelsen, ``Open cell library in 15nm freepdk technology,'' in
  \emph{Proceedings of the 2015 Symposium on International Symposium on
  Physical Design}, ser. ISPD '15.\hskip 1em plus 0.5em minus 0.4em\relax New
  York, NY, USA: ACM, 2015, pp. 171--178. [Online]. Available:
  \url{http://doi.acm.org/10.1145/2717764.2717783}
\BIBentrySTDinterwordspacing

\bibitem{SHWW14:ICCAD}
E.~Schneider, S.~Holst, X.~Wen, and H.-J. Wunderlich, ``Data-parallel
  simulation for fast and accurate timing validation of cmos circuits,'' in
  \emph{2014 IEEE/ACM International Conference on Computer-Aided Design
  (ICCAD)}, 2014, pp. 17--23.

\bibitem{strang2014differential}
G.~Strang, \emph{Differential equations and linear algebra}.\hskip 1em plus
  0.5em minus 0.4em\relax Wellesley-Cambridge Press Wellesley, 2014.

\bibitem{MSIM:FLI}
{Mentor Graphics Corporation}, \emph{Foreign Language Interface Manual}, 2016,
  {Software Version 10.5c}.

\end{thebibliography}
\end{document}